\documentclass{aip-cp}

\usepackage[numbers]{natbib}
\usepackage{rotating}
\usepackage{graphicx}
\usepackage{savesym}
\savesymbol{iint}
\savesymbol{iiint}
\savesymbol{iiiint}
\savesymbol{idotsint}
\usepackage{amsmath}
\restoresymbol{AMS}{iint}
\restoresymbol{AMS}{iiint}
\restoresymbol{AMS}{iiiint}
\restoresymbol{AMS}{idotsint}

\begin{document}

\title{Non-Equilibrium Many-Body Dynamics Following A Quantum Quench}

\author[aff1]{Manan Vyas\corref{cor1}}

\affil[aff1]{Instituto de Ciencias F{\'i}sicas, Universidad Nacional Aut{\'o}noma de M{\'e}xico, 62210 Cuernavaca, M{\'e}xico}
\corresp[cor1]{Corresponding author: manan@icf.unam.mx}

\maketitle

\begin{abstract}
We study analytically and numerically the non-equilibrium dynamics of an isolated interacting many-body quantum system following a random quench. We model the system Hamiltonian by Embedded Gaussian Orthogonal Ensemble (EGOE) of random matrices with one plus few-body interactions for fermions. EGOE are paradigmatic models to study the crossover from integrability to chaos in interacting many-body quantum systems. We obtain a generic formulation, based on spectral variances, for describing relaxation dynamics of survival probabilities as a function of rank of interactions. Our analytical results are in good agreement with numerics. 
\end{abstract}

\section{INTRODUCTION}

Understanding the non-equilibrium dynamics of interacting many-body quantum systems is fundamental for various fields of physics \cite{Ma-98, Ma-16, Bo-16, Mu-09}. Unitary quantum dynamics is investigated experimentally using cold atoms, ion traps and nuclear magnetic resonance \cite{Ka-16, Sm-16, Ga-16, We-16}. Statistical models describing realistic interacting many-body systems, such as nuclei, ultracold atoms, quantum dots etc, are of extensive interest to understand univeral features arising due to interactions.

We model complex quantum systems by realistic random matrices, that incorporate few-body nature of interactions, known as EGOE($k$) with rank of interaction $k$ much smaller than the number of fermions in the system. For more details about the known results for EGOE($k$), refer to \cite{Br-81, Be-03,Ko-14}. EGOE are generic though analytically difficult to deal with in comparison to canonical GOE. EGOE are paradigmatic models to study many-body quantum chaos \cite{Pa-07}. EGOE($k$) have been recently used to study transport properties in disordered systems \cite{Ma-15,Be-16}. 

We analyze relaxation dynamics of an isolated many-body quantum system following a random interaction quench. The system is initially prepared in one of the eigenstates of unperturbed mean-field Hamiltonian $H_0$. Dynamics begins with instantaneous change of quench strength $\lambda$ from zero resulting in final Hamiltonian $H = H_0 + \lambda \; V(k)$ with eigenvalues $E$. Here, $V(k)$ is the random $k$-body perturbation [EGOE($k$)]. In fact, the final Hamiltonian $H$ is also known in the literature as EGOE(1+2) for $k=2$. So, we will also use the notation EGOE(1+$k$) to denote the final Hamiltonian $H$. We focus on survival probability to characterize system evolution. We show that relaxation dynamics is generic and depends on the variance (spread in energy) of the initial state.

This paper is organized as follows: in Section II we define EGOE with $k$-body interactions. In Section III, we study the semi-circle to Gaussian transition in eigenvalue densities and briefly discuss the formulaes for the second and fourth moments for the eigenvalue densities. Section IV gives the generic formulation and numerical results for relaxation dynamics of survival probabilities. Finally, Section V gives conclusions.

\section{EMBEDDED GAUSSIAN ORTHOGONAL ENSEMBLE: DEFINITION}

Consider a system of $m$ identical (spinless) fermions distributed in $N$ degenerate single particle (sp) levels with $k$-body interactions ($k \leq m$). The embedding algebra for EGOE($k$) is $SU(N)$. These ensembles are defined by three parameters $(N,m,k)$ and the random $k$-body Hamiltonian in second quantized form is,
\begin{equation}
V(k) = \displaystyle\sum_{\alpha,\;\beta} \; v^{\alpha,\;\beta}_k \; \alpha^\dagger(k) \; \beta(k) \;.
\label{eq-1}
\end{equation} 
Here, $\alpha^\dagger(k)$ and $\beta(k)$ respectively are $k$-particle creation and annihilation operators for fermions. They obey the usual anti-commutation relations. In 
Equation \eqref{eq-1}, $v^{\alpha,\;\beta}_k$ are random distributed independent Gaussian variables with zero mean and variance
\begin{equation}
{\overline{v^{\alpha,\;\beta}_k \; v^{\alpha^\prime,\;\beta^\prime}_k}} = v^2 \; \left( {\delta_{\alpha,\;\beta^\prime}} {\delta_{\alpha^\prime,\;\beta}} + {\delta_{\alpha,\;\alpha^\prime}} {\delta_{\beta^\prime,\;\beta}} \right)
\label{eq-2}
\end{equation} 
Here, the bar denotes ensemble average and we set $v=1$ without loss of generality. In other words, $v^{\alpha,\;\beta}_k$ is chosen to be a $\binom{N}{k}$ dimensional GOE in $k$-particle spaces.

Each possible distribution of $m$ fermions in the $N$ sp levels (with $N > m$) generates a configuration or a basis state. Distributing the $m$ fermions in all possible ways in $N$ levels generates the $d(N,m)=\binom{N}{m}$ dimensional Hilbert space or basis space. This is similar to distributing $m$ particles in $N$ boxes with the conditions that the occupancy of each box can be either zero or one and the total number of occupied boxes equals $m$. Given the sp states $| \nu_i\rangle$, $i=1,\;2,\ldots,\;N$, the action of the Hamiltonian operator $V(k)$ defined by Equation \eqref{eq-1} on the basis states $| \nu_1 \nu_1 \ldots \nu_m \rangle$ generates the EGOE($k$) ensemble in $m$-particle spaces. Note that, the notation $| \nu_1 \nu_1 \ldots \nu_m \rangle$ denotes the levels occupied by the $m$ spinless fermions. By construction, the case $k = m$ is identical to a canonical GOE. Unlike GOE, EGOE($k < m$) incorporates few-body nature of interactions that results in correlations between matrix elements of $V(k)$ and many of them will be zero due to $k$-body selection rules.  

\section{SEMI-CIRCLE TO GAUSSIAN TRANSITION IN EIGENVALUE DENSITY}

Binary correlation theory for deriving ensemble averaged moments generated by EGOE($k$) in the dilute limit ($N \to \infty,\; m \to \infty,\; m/N\to 0$) is given by Mon and French \cite{Mo-75} and has been extended for deriving moments defined over two-orbits  \cite{Ma-15a}. We briefly mention the required results for the ensemble averaged variances and kurtosis for EGOE($k$). For more details, refer to \cite{Ma-12,Ma-15a}.

Incorporating the finite-$N$ corrections and using ${\overline{(v^{\alpha,\;\beta}_k)^2}} =\lambda^2$, the ensemble averaged variances are given by,
\begin{equation}
\sigma^2_{V(k)} = {\overline{\langle V^2(k) \rangle^m}} = \lambda^2 \; \displaystyle\binom{m}{k} \left[ \displaystyle\binom{N-m+k}{k} +1 \right]
\label{eq-3}
\end{equation} 
Similarly, the expression for fourth order moment is,
\begin{eqnarray}
{\overline{\langle V^4(k) \rangle^m}} & = & \lambda^4 \; \left[ 2 \; \displaystyle\binom{m}{k}^2 \left\{ \displaystyle\binom{N-m+k}{k} +1 \right\}^2 + \displaystyle\sum_{s = 0}^k \; \displaystyle\binom{m-s}{k-s}^2 \; \displaystyle\binom{N-m+k-s}{k} \; \displaystyle\binom{m-s}{k}\; \displaystyle\binom{N-m}{s} \; \displaystyle\binom{m}{s} \right.
 \\ 
& \times & \left. \displaystyle\binom{N+1}{s} \; \displaystyle\frac{N-2s+1}{N-s+1}\; \displaystyle\binom{N-s}{k}^{-1} \; \displaystyle\binom{k}{s}^{-1}  \right] \nonumber 
\label{eq-4}
\end{eqnarray}
Finally, kurtosis in the strict $N \to \infty$ limit is,
\begin{eqnarray}
\gamma_2 & = &  {\overline{\langle V^4(k) \rangle^m}} \; \left[  {\overline{\langle V^2(k) \rangle^m}} \right]^{-2} -1 \nonumber \\
& \to & \displaystyle\frac{-k^2}{m} + O(m^{-2}) 
\label{eq-5}
\end{eqnarray} 

\noindent For EGOE($k$), eigenvalue densities ${\overline{\rho(E)}} = \langle \delta(V - E)\rangle^m$ exhibit semi-circle to Gaussian transition with increasing rank of interaction $k$ for a fixed-$m$. The rate at which the transition occurs can be measured by kurtosis. The value $|\gamma_2| > 0.3$ characterizes departure from Gaussian. The transition from semi-circle to Gaussian can also be generated for a fixed-$k$ by increasing number of fermions $m$. The eigenvalue densities approach Gaussian form for large $m$ (with fixed-$k$) and small $k$ (with fixed-$m$) as clearly demonstrated by Equation \eqref{eq-5}.  

\begin{figure}[ht]
  \centerline{\includegraphics[width=5.5in]{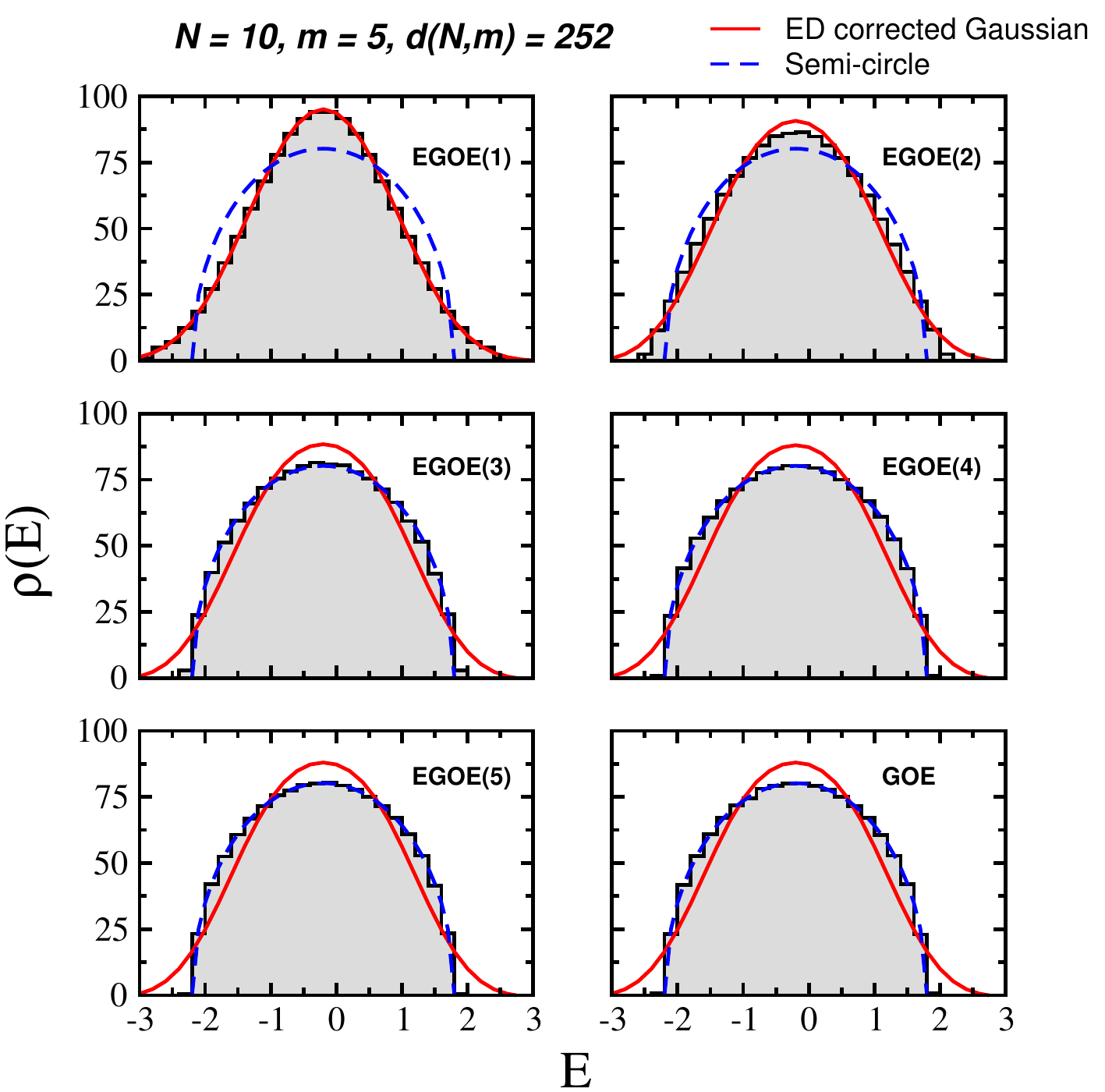}}
  \caption{Ensemble averaged eigenvalue densities $\rho(E)$ for a 1000 member EGOE($k$) with $k=1-5$ and GOE. The continuous curves are ED corrected Gaussians [Equation \eqref{eq-6}] and the dashed curves are semi-circle. In the plots, the histograms are normalized to dimension $d(N,m)$ and the energies $E$ are normalized energies.}
\label{dos}
\end{figure}

Figure \eqref{dos} shows ensemble averaged eigenvalue densities for a 1000 member EGOE($k$) for 5 fermions distributed in 10 sp levels as a function of rank of interactions $k = 1, \;2,\ldots,\;5$. For comparison, we also show the eigenvalue density for canonical GOE of same dimension. The numerical histograms are compared with semi-circle and Edgeworth (ED) corrected Gaussian (which incorporates corrections to Gaussian due to skewness $\gamma_1$ and kurtosis $\gamma_2$) \cite{St-87},
\begin{equation}
\rho_{ED}(E) = \displaystyle\frac{1}{\sqrt{2\pi}} \; \exp\left( - \displaystyle\frac{E^2}{2} \right) \left\{  1 + \displaystyle\frac{\gamma_1}{6} He_3(E) + \displaystyle\frac{\gamma_2}{24} He_4(E) + \displaystyle\frac{\gamma_1^2}{72} He_6(E) \right\} \;.
\label{eq-6}
\end{equation} 
Here, $He$ are the Hermite polynomials: $He_3(x) = x^3-3x$, $He_4(x) = x^4-6x^2+3$, and $He_6(x) = x^6-15x^4+45x^2-15$. Also, $E$ are the normalized energies, i.e. centroids are zero and variances are unity. 

As seen from the figure, the eigenvalue density is Gaussian for $k = 1$ and $2$ and becomes semi-circular at $k = 3$. Also, we see that the eigenvalue densities for EGOE(5) and GOE are identical. Thus, eigenvalue densities make a transition from Gaussian to semi-circle with increasing $k$ and is well explained by Equation \eqref{eq-5}. Also, the numerical values of variances and kurtosis for the eigenvalue densities match with the values obtained from Equations \eqref{eq-3} and \eqref{eq-5} respectively.

\section{DYNAMICS: SURVIVAL PROBABILITY}

\begin{figure}[t]
  \centerline{\includegraphics[width=4.95in]{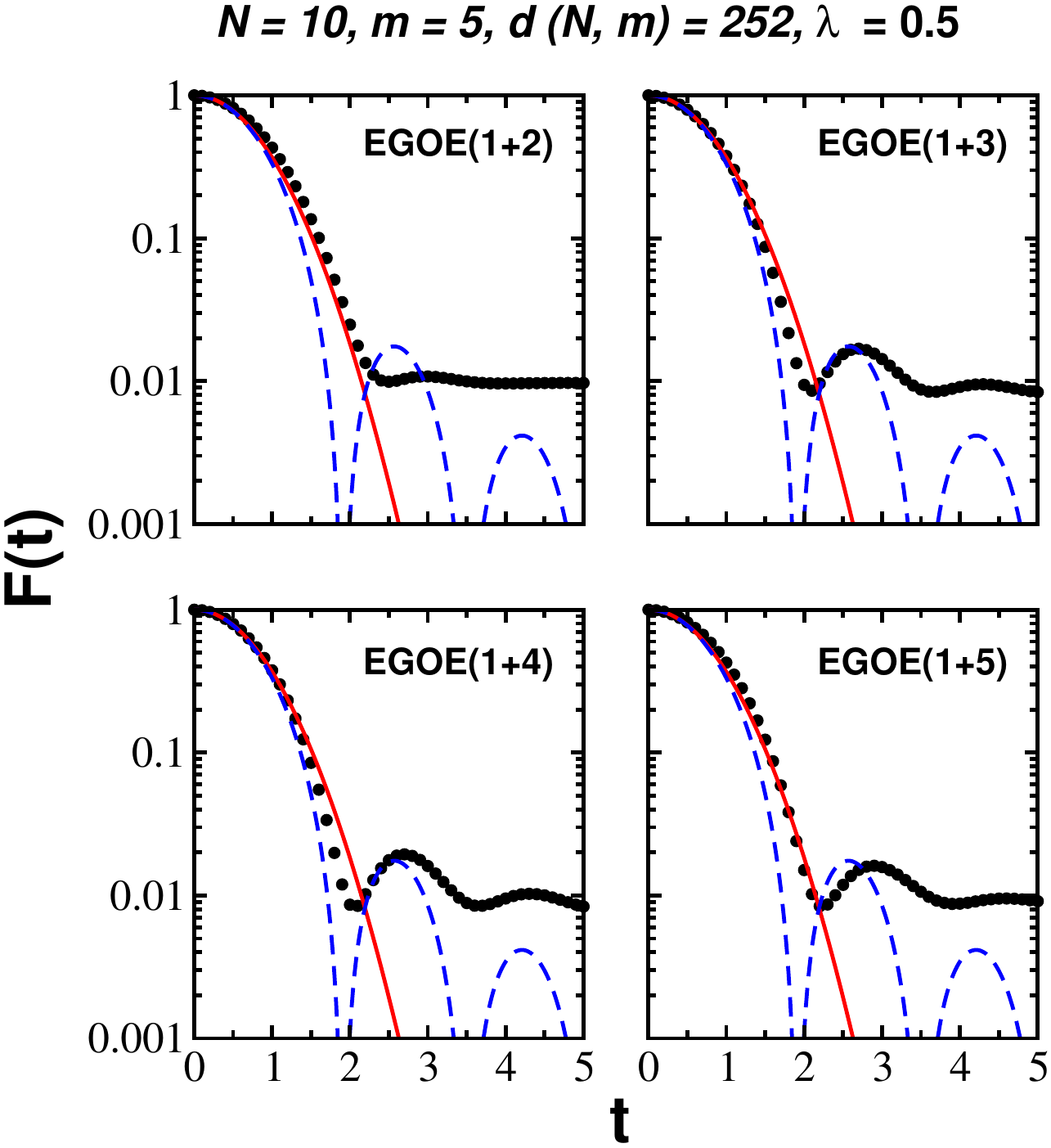}}
  \caption{Ensemble averaged survival probabilities $F(t)$ for a 1000 member $H$ ensemble as a function of rank of interaction $k$. Numerical results (solid circles) are compared with Gaussian (solid curves) and Bessel function (dashed curves) behavior. The corresponding ensemble averaged variances $\sigma_0^2(k)$ are $0.734$, $0.878$, $0.889$ and $0.761$ for $k=2-5$ respectively.}
\label{fid-l0.5}
\end{figure}

Initially, the quantum system (time-reversal and rotationally invariant) is in the eigenstate $|ini\rangle = \psi(0)$ of unperturbed mean-field Hamiltonian $H_0 = \sum_{i} \epsilon_i \; \hat{n}_i$. Note that the $\epsilon_i$ are sp energies with unit average level spacing and $\hat{n}_i$ are number operators acting on the sp levels $i = 1$, 2, $\ldots, \; N$. We choose $\epsilon_i = i + 1/i$ to avoid degeneracies in the many-particle spectrum. 

Dynamics begins by quenching the Hamiltonian $H_0$ to a new final Hamiltonian $H$,
\begin{equation}
H = H_0 + \lambda \; V(k) \;,
\label{eq-7}
\end{equation}
with $V(k)$ given by Equation \eqref{eq-1}. Note that $\lambda$ is the strength of the random $k$-body perturbation. The initial state $\psi(0)$ evolves unitarily under quench $V(k)$ after time $t$ to $\psi(t) = \exp(-iHt) \; \psi(0)$. Survival probability $F(t)$ is the probability for finding the system still in initial state $\psi(0)$ after time $t$,
\begin{equation}
F(t)  = |A(t)|^2 = |\langle \psi(0) |e^{-iHt} |\psi(0) \rangle |^2 \;,
\label{eq-8}
\end{equation}
where $A(t)$ is the survival amplitude. By projecting the initial state $\psi(0)$ on the eigenstates $|E\rangle$ of $H$, $A(t) = \int |C_0^E|^2 \; \rho(E) \; e^{-iEt} dE$, which is the Fourier transform of strength functions or the LDOS. Strength functions give the spreading of the basis states over the eigenstates and thus, fourier transform of the strength functions in energy gives the survival amplitude and thus, survival probability.  Here, $C_0^E = \langle E \;|\; \psi(0) \rangle$ are the overlaps.

Following previous works \cite{Sa-01,Ko-01, Ma-11}, we know that EGOE(1+2) exhibits three chaos markers ($\lambda_1$ \textless $\lambda_2$ \textless $\lambda_3$) as a function of perturbation strength $\lambda$. For $\lambda \sim \lambda_3$, strength functions are Gaussians and therefore, survival probability follows Gaussian law: 
\begin{equation}
F(t) = \exp[-\sigma_0^2(k) \; t^2],
\label{eq-8}
\end{equation} 
with $\sigma_0^2(k)$ being the (ensemble averaged) variance of the initial state for a given $k$. For EGOE(1+$k$), $\sigma_0^2(k)$ is given by,
\begin{equation}
\sigma_0^2(k) =  \displaystyle\frac{{\sigma^2_{V(k)}}}{{\sigma^2_{H}}} \;.
\label{eq-9}
\end{equation}
Note that, $\sigma^2_{H} = \sigma^2_{H_0} + \sigma^2_{V(k)}$ is the ensemble averaged variance for the final Hamiltonian $H$ with $\sigma^2_{V(k)}$ given by Equation \eqref{eq-3}.
Similarly, for semicircular strength functions, the survival probability is given by 
\begin{equation}
F(t) = \displaystyle\frac{[\mathcal{J}_1(2\; \sigma_0(k)\; t)]^2}{\sigma_0^2(k) \; t^2} \;, 
\label{eq-10}
\end{equation}
where $\mathcal{J}_1$ is the Bessel function of first kind \cite{Ma-14}.

For a 1000 member $H$ ensemble with $N = 10$, $m = 5$ and $\lambda = 0.5$, survival probabilities are shown in Figure \eqref{fid-l0.5} as a function of rank of interaction $k$ varying from 2 to 5. We choose initial states $\psi(0)$ in an energy window of size $\delta = 0.01$ around center of the spectrum. For each member of the ensemble, the eigenvalues $E$ and the initial state energies are normalized and then the survival probabilities are summed over the initial states. It is important to note that we carry out ensemble average over the initial states as well. Relaxation dynamics of the ensemble averaged survival probabilities in Figure \eqref{fid-l0.5} are compared with Gaussian [Equation \eqref{eq-8}] and Bessel function [Equation \eqref{eq-10}] behavior; the values of $\sigma_0^2(k)$ are computed for these using Equation \eqref{eq-9}. Agreement 
between numerical results and Gaussian law is excellent for short times, in general. However, the agreement gets better with increasing $k$. This is expected as $\lambda_3$ is smallest for largest $k$. We also see clear departure from the Bessel function behavior for all $k$ values. We have also verified that survival probability follows Bessel function law for EGOE($k=m$) for higher dimensional systems. Thus, our analytical formulation explains the relaxation dynamics of survival probability as a function of rank of random interaction. 

\section{CONCLUSIONS}

Generic features of survival probabiities explored in the context of random $k$-body interactions is an important step in establishment of general description of non-equilibrium dynamics of interacting quantum many-body systems. Our formulation gives analytical description of relaxation dynamics of complex quantum systems in the absence of details of system dynamics. Further detailed analysis will be reported in a separate future publication.

\section{ACKNOWLEDGMENTS}
Thanks are due to Lea Santos and Luis Benet for many useful discussions.
Author acknowledges supercomputing facility LANCAD-UNAM-DGTIC-330 
and financial support from UNAM/DGAPA/PAPIIT research grant IA104617 and CONACyT research grant 219993.


\nocite{*}
\bibliographystyle{aipnum-cp}%
\bibliography{manan-pap-NMP}%

\end{document}